\documentclass[aip,jap,preprint]{revtex4-1}

\usepackage{graphicx}
\usepackage{dcolumn}
\usepackage{bm}

\usepackage[T1]{fontenc}
\usepackage{mathptmx}

\usepackage{braket}
\usepackage{graphicx}
\usepackage{xcolor}
\usepackage{ulem}

\draft 

\begin{document}

\preprint{The following article has been accepted by Journal of Applied Physics.} \preprint{After it is published, it will be found at https://doi.org/10.1063/5.0016069.}

\title[]{Asymmetric versus symmetric $\rm{HgTe/Cd_{x}Hg_{1-x}Te}$ double quantum wells: Band gap tuning without electric field} 



\author{Du\v{s}an B. Topalovi\'c}
\email[]{dusan.topalovic@vin.bg.ac.rs}
\affiliation{School of Electrical Engineering, University of Belgrade, P. O. Box 35-54, 11120 Belgrade, Serbia}
\affiliation{Department of Radiation and Environmental Protection, Vin\v{c}a Institute of Nuclear Sciences-National Institute of the Republic of Serbia, University of Belgrade, P. O. Box 522, 11001 Belgrade, Serbia}

\author{Vladimir V. Arsoski}
\email[]{vladimir.arsoski@etf.bg.ac.rs}
\affiliation{School of Electrical Engineering, University of Belgrade, P. O. Box 35-54, 11120 Belgrade, Serbia}

\author{Milan \v{Z}. Tadi\'c}
\email[]{milan.tadic@etf.bg.ac.rs}
\affiliation{School of Electrical Engineering, University of Belgrade, P. O. Box 35-54, 11120 Belgrade, Serbia}

\author{Fran\c{c}ois M. Peeters}
\email[]{francois.peeters@uantwerpen.be}
\affiliation{Department of Physics, University of Antwerp, Groenenborgerlaan 171, B-2020 Antwerp, Belgium}
\affiliation{School of Physics and Astronomy and Yunnan Key Laboratory for Quantum Information, Yunnan University, Kunming 650091, China}



\begin{abstract}
We investigate the electron states in double asymmetric $\rm{HgTe/Cd_{x}Hg_{1-x}Te}$ quantum wells grown along the $[001]$ direction. The subbands are computed by means of the envelope function approximation applied to the 8-band Kane $\bf{k}\cdot\bf{p}$ model. The asymmetry of the confining potential of the double quantum wells results in a gap opening which is absent in the symmetric system where it can only be induced by an applied electric field. The band gap and the subbands are affected by spin-orbit coupling which is a consequence of the asymmetry of the confining potential. The electron-like and hole-like states are mainly confined in different quantum wells, and the enhanced hybridization between them opens a spin-dependent hybridization gap at a finite in-plane wavevector. We show that both the ratio of the widths of the two quantum wells and the mole fraction of the $\rm{Cd_{x}Hg_{1-x}Te}$ barrier control both the energy gap between the hole-like states and the hybridization gap. The energy subbands are shown to exhibit inverted ordering, and therefore a nontrivial topological phase could emerge in the system.
\end{abstract}


\maketitle 

\section{Introduction}
\label{intro}
New phases of matter have recently been discovered \cite{Hasan2010,Wen1990}, that have distinct topological order for which the Landau theory of symmetry breaking is not applicable. Early examples are the integer and the fractional quantum Hall effects \cite{Thouless1982,Tsui1982}. Recently, a large class of materials characterized by specific conductive edge states named topological insulators (TI's), which have emergent applications in electronics and photonics, were discovered \cite{Liu2014,Zhu2013,Ezawa2013,Cho2011,Xiu2011}. They have peculiar edge states which are studied in Ref. \onlinecite{Kalmeyer1987}. It was found that some of these chiral spin states have the same symmetry, yet they differ by their topological invariance \cite{Thouless1982,Nakahara2003,Berry1984,Kane2005a}.

The interior of a TI sample is insulating and the electric current passes only along the sample's surface \cite{Shen2012}. Such TI has {\it helical edge states} which are located in the energy gap. These states are protected by time-reversal symmetry, which do not exist in a normal insulator (NI). The TI samples made by tailoring two-dimensional (2D) materials exhibit the quantum spin Hall effect, which is a peculiar TI effect. TI effects were theoretically predicted for graphene \cite{Kane2005b}, where the quantized spin Hall conductance was explained by the Haldane model \cite{Haldane1988}. The spin-orbit interaction in a system consisting of light carbon atoms is too small to give rise to an energy gap \cite{Wang2015}. Hence, the attention has focused to materials made of heavier atoms which are influenced by relativistic effects. Bernevig et al. devised a simple effective model which showed that a quantum spin Hall state exists in $\rm{HgTe/Cd_{x}Hg_{1-x}Te}$ quantum wells \cite{Bernevig2006}. It was found that the conduction and the valence bands become inverted for well thicknesses above a critical value $d_{c}=6.3$ $\rm{nm}$ \cite{Bernevig2006}.

The theoretical predictions were subsequently confirmed in a beautiful experiment of K\"onig et al. \cite{Konig2007}. Symmetric double quantum wells (SDQW's) based on $\rm{HgTe/Cd_{x}Hg_{1-x}Te}$ and $\rm{InAs/GaSb/AlSb}$ were also explored \cite{Michetti2012,Liu2008}. The presence of topologically non-trivial phases was demonstrated, with the possibility of tuning the transition to the topological state by means of varying an external electric field. Moreover, a recent study showed that the $\rm{HgTe/Cd_{x}Hg_{1-x}Te}$ SDQW possesses the 3/2 pseudospin degree of freedom, and thus interesting effects due to spin coupling between adjacent HgTe layers where predicted \cite{Krishtopenko2016}. These results are similar to both the physics of pseudospin in bilayer graphene (BG) without valley degeneracy \cite{Michetti2012} and some recently analyzed type-II and broken-gap quantum wells \cite{Li2009,Xu2010,Hu2016}.

Motivated by the studies of topological effects in systems of quantum wells, we investigate hereafter the electronic properties of {\it asymmetric} $\rm{HgTe/Cd_{x}Hg_{1-x}Te}$ double quantum wells (ADQW's). Our aim was to explore the effects due to the inversion between the $s$-like valence band with $\mathrm{\Gamma}_{6}$ symmetry and the $p$-like conduction band with $\mathrm{\Gamma}_{8}$ symmetry. In section \ref{sec:2} the $\bf{k}\cdot\bf{p}$ model based on the envelope function theory and a procedure for solving the eigenvalue problem by expanding the envelope functions into a complete basis set of plane waves are presented. Also, the criterion to unequivocally classify the subbands states in an ADQW system is formulated. The results of our calculations are presented and discussed in section \ref{sec:3}, where our findings are summarized and a concise conclusion is given in the final section.

\section{Theoretical model}
\label{sec:2}
\begin{figure}
	\includegraphics{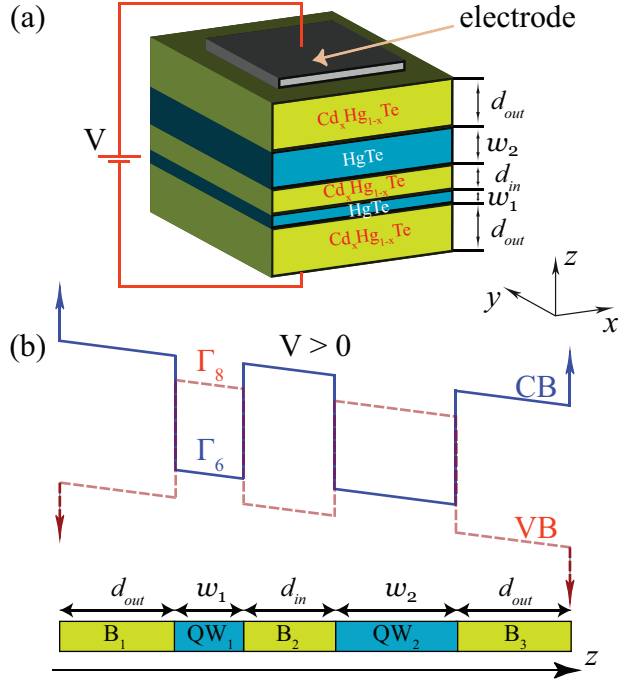}
	\caption{(a) A 3D schematic view of the analyzed $\rm{HgTe/Cd_{x}Hg_{1-x}Te}$ ADQW structure. The $z$ axis is assumed to be directed perpendicular to the layers and oriented along the $[001]$ crystallographic direction. The HgTe wells, whose widths are denoted by $w_{1}$ and $w_{2}$, are separated by a barrier whose thickness is denoted by $d_{in}$, and are surrounded by barriers of the same composition and thickness $d_{out}$. (b) The band diagram of the lowest conduction band (denoted by CB) and the highest valence band (denoted by VB) when voltage $V$ is applied across the ADQW. The bands in the HgTe layers are inverted at $k_{\parallel}=0$, with the $p$-like $\mathrm{\Gamma}_{8}$ band having higher energy than the $s$-like $\mathrm{\Gamma}_{6}$ band.}
	\label{fig:1}
\end{figure}

Figure \ref{fig:1} shows a schematic view of the analyzed double quantum wells. There are two HgTe quantum wells separated by a $\rm{Cd_{x}Hg_{1-x}Te}$ barrier, and the $z$-axis points along the growth direction. Widths of the quantum wells are denoted by $w_{1}$ and $w_{2}$, whereas the interwell barrier thickness is denoted by $d_{in}$. In our model, the two wells are assumed to be surrounded by two barriers of equal width, denoted by $d_{out}$, whose composition is assumed to be equal to the composition of the interwell barrier. Hence, the considered structure has finite width.

The electronic structure of the $\rm{HgTe/Cd_{x}Hg_{1-x}Te}$ double quantum well (DQW) is extracted from the 8-band $\bf{k}\cdot\bf{p}$ model \cite{Kane1957,Novik2005}
\begin{eqnarray}
	H{\rm{\Xi}}({\mathbf{r}})=E{\rm{\Xi}}({\mathbf{r}}),\label{eq:1}
\end{eqnarray}
which is a set of 8 coupled differential equations. Here, $H$ is the multiband Hamiltonian, and $\Xi$ denotes the envelope-function spinor composed of the envelope functions $\rm{\chi}$$_{n}({\mathbf{r}})$, $n\in\{1,2,...,8\}$, which correspond to the zone-center Bloch states forming the basis. We note that the same basis functions are used for both the well and the barriers. Furthermore, because of translational symmetry in the layers plane, the variables $x$ and $y$ can be separated from $z$,
\begin{eqnarray}
	{\rm{\Xi}}_{\bf k_\parallel}(x,y,z)={\rm{exp}}\left[i(k_{x}x+k_{y}y)\right]\left[\chi_{1,(k_x,k_y)}(z)\mskip 10mu\chi_{2,(k_x,k_y)}(z)\mskip 10mu...\mskip 10mu\chi_{n,(k_x,k_y)}(z)\right]^{T},\label{eq:3}
\end{eqnarray}
where $k_{x}$ and $k_{y}$ are the components of the in-plane electron wave vector. A more detailed description of the Hamiltonian is given in the Appendix. The eigenvalue problem is solved by expanding the envelope functions into a complete basis set composed of plane waves \cite{Cukaric2016}
\begin{eqnarray}
	\chi_{s,(k_x,k_y)}(z)=\frac{1}{\sqrt{L_{sc}}} \sum_{p=-N}^{N} c_{s,p}(k_x,k_y)\cdot\exp\left({\frac{i2p\pi z}{{L_{sc}}}}\right),\label{eq:4}
\end{eqnarray}
where $s\in\{1,2,...,n\}$, $c_{s,p}(k_x,k_y)$ are the expansion coefficients, and $L_{sc}=w_1+w_2+$$d_{in}+2d_{out}$. The basis is limited to the first Brillouin zone, i.e. $2\pi |N|/L_{sc}\leq2\pi/a_{0}$, where $a_{0}$ is the lattice constant. The Hamiltonian matrix is of the order of $8\times(2N+1)$. An external electric field $F_{el}$ is taken into account by adding the matrix elements of $\varphi(z)=e\:{\mathbf F}_{el}\cdot {\mathbf r}= eF_{el}z=eV(z)$ to the diagonal of the Hamilton matrix. The electric field modifies the band edges as shown in Figure {\ref{fig:1}}(b).

Since the bands are mixed in the adopted ${\bf k}\cdot{\bf p}$ theory, it became necessary to classify the states. It was done by computing a probability which corresponds to the Kramers degenerate pair of the basis states \cite{Krishtopenko2016}:
\begin{eqnarray}
	p_{S}(k_{x},k_{y})=\sum_{p=-N}^{N}\sum_{s\in S}|c_{s,p}(k_{x},k_{y})|^2,\label{eq:5}
\end{eqnarray}
where $S=\{1,2\}$ for the conduction-band states probability $p_{cb}(k_{x},k_{y})$, $S=\{3,6\}$ for the heavy-hole states  probability $p_{hh}(k_{x},k_{y})$, $S=\{4,5\}$ for the light-hole states probability $p_{lh}(k_{x},k_{y})$, and $S=\{7,8\}$ for the spin-orbit split-off states probability $p_{so}(k_{x},k_{y})$. $p_{cb}$, $p_{hh}$, $p_{lh}$, and $p_{so}$ are used to show how the $\ket{\mathrm{\Gamma}_{6},\pm1/2}$, $\ket{\mathrm{\Gamma}_{8},\pm3/2}$, $\ket{\mathrm{\Gamma}_{8},\pm1/2}$, and $\ket{\mathrm{\Gamma}_{7},\pm1/2}$ zone-center states contribute to the DQW states. The states are classified to be $hh$-like if $p_{hh}>p_{lh}+p_{cb}+p_{so}$ at $k_{\parallel}=0$. On the other hand, if $p_{lh}+p_{cb}+p_{so}>p_{hh}$ the subband state is classified as either the conduction-band-like, light-hole-like, or spin-orbit-like for the case of the largest $p_{cb}$, $p_{lh}$, or $p_{so}$, respectively \cite{Krishtopenko2016}.

\section{Results and discussion}
\label{sec:3}

The parameters of the band structure at $T=0$ that we used for the calculations are taken from Ref. \onlinecite{Sengupta2013}. For these parameters there is an excellent agreement of numerical\cite{Sengupta2013} and experimental \cite{Brune2012} results in CdTe/HgTe/CdTe quantum well heterostructure. We explore how the asymmetry of the confining potential affects the electronic structure of the $\rm{HgTe/Cd_{x}Hg_{1-x}Te}$ double quantum wells (DQW). The obtained results will be contrasted with the electronic structure of the SDQW's which were previously modelled in Refs.~\onlinecite{Michetti2012,Krishtopenko2016}. The values of the barrier thickness $d_{out}$, the widths of the HgTe wells $w_{1}$, and $w_{2}$, and the $\rm{HgTe/Cd_{x}Hg_{1-x}Te}$ spacer thickness $d_{in}$ adopted for our calculations are displayed in Table \ref{tab:1} for both the SDQW  and ADQW. The mole fraction of CdTe in the $\rm{Cd_{x}Hg_{1-x}Te}$ alloy is taken to be equal to $x=0.7$.

\begin{table}
	\caption{The values of the characteristic dimensions of a DQW: $d_{out}$, $w_{1}$, $w_{2}$, and $d_{in}$ (see Figure \ref{fig:1}) used in our numerical calculations.}
	\label{tab:1}
	\begin{ruledtabular}
		\begin{tabular}{cccccc}
			& Type & $d_{out} \rm{(nm)}$ & $w_{1} \rm{(nm)}$ & $w_{2} \rm{(nm)}$ & $d_{in}\rm{(nm)}$  \\
			\hline
			SDQW & Symmetric & 30 & 6 & 6 & 2.5 \\
			ADQW & Asymmetric & 30 & 6 & 7.8 & 2.5 \\
		\end{tabular}
	\end{ruledtabular}		
\end{table}

\begin{figure}
	\includegraphics{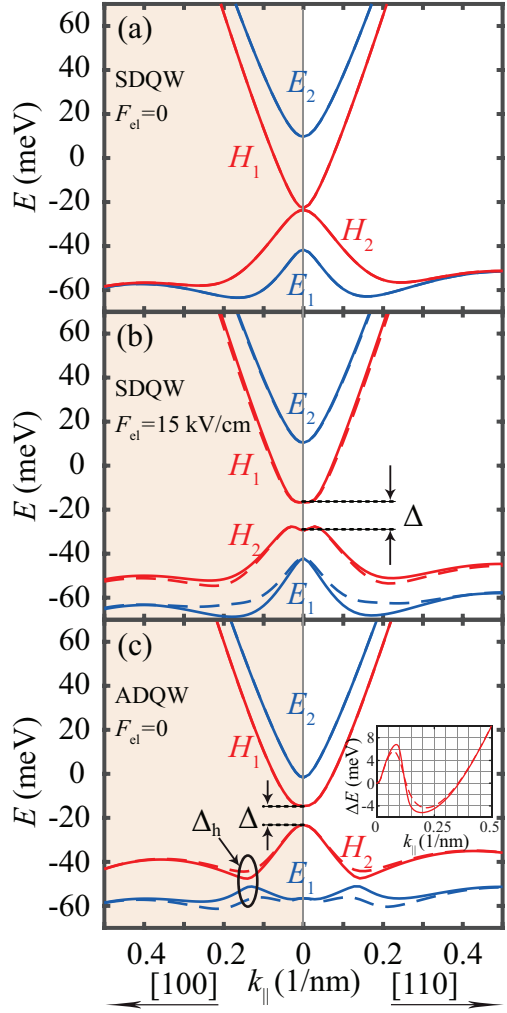}
	\caption{The subband dispersion relations in: (a) SDQW in the absence of an electric field, (b) SDQW for electric field of $15$ kV/cm, and (c) ADQW in the absence of an electric field. Note that in (a) there is no band gap, an indirect band gap $\Delta$ appears in (b) when a perpendicular electric field is applied, and in (c) ADQW exhibits both a direct gap $\Delta$ and a hybridization gap denoted by $\mathrm{\Delta_h}$ even in the absence of an electric field. The inset shows the difference between  the energy eigenvalues along the [110] and [100] directions for $H_2$ subband. The values of the geometric dimensions of the structures and the band-structure parameters are given in Table \ref{tab:1} and in the text. The dispersions along the [100] ([110]) direction are shown in left (right) part of the figure, and the energies of the spin-up (spin-down) states are displayed by solid (dashed) lines.}
	\label{fig:2}
\end{figure}

The quantum wells in the SDQW are assumed to be 6 nm wide. In order to enhance tunneling between the quantum wells, the wells are separated by a thin barrier which is only $d_{in}=2.5$ nm thick. The SDQW contains a metal phase whose properties are similar to the BG \cite{Krishtopenko2016}. The main feature of the BG phase is the existence of a zero-energy gap which might be opened by means of the gate voltage \cite{Min2007}. However, the major advantage of the HgTe DQW over BG is the existence of topologically protected edge states in the gap. In the $\rm{HgTe/Cd_{x}Hg_{1-x}Te}$ SDQW's there exists a specific band ordering in the BG-like phase. The bands were classified by the properties of the wave functions (see Eq. (\ref{eq:5})) at $k_\parallel=0$ \cite{Bernevig2006,Xu2010}. Four subbands are of interest here. Two of them, which are labeled by $E_{1}$ and $E_{2}$, are electron-like, whereas the other two, denoted by $H_{1}$ and $H_{2}$, are hole-like. The energy dispersion relations of these subbands are shown in Figure \ref{fig:2}. Here, the dispersion relations shown in the left (right) panel correspond to the $[100]$ ($[110]$) direction, and the spin-up (spin-down) subbands are displayed by solid (dashed) lines. The spin of a subband is determined from the wave function properties at $k_\parallel=0$ \cite{Bernevig2006,Xu2010}.

The total angular momentum of the $H_{1}$ and $H_{2}$ states is $j=3/2$, and as Figure \ref{fig:2}(a) shows, in the absence of an external electric field they are degenerate at $k_\parallel=0$. We note that the dispersion relations of all the subbands are almost isotropic in the $(k_x,k_y)$ plane, and they are parabolic in close vicinity of the Brillouin-zone center. We noticed that out of all the shown subbands in the ADQW $H_2$ exhibits the highest in-plane anisotropy. The difference between the energy eigenvalues of these subbands computed for the [110] and [100] directions, $\Delta E = E_{H_2}^{[110]}-E_{H_2}^{[100]}$, can be as large as 6 meV, for even not so larger $k_{\parallel}$, of the order of 0.1 ${\rm nm}^{-1}$, which is shown in the inset in Figure \ref{fig:2}(c). The $E_{2}$ subband is evidently above the $H_{1}$ subband, but $E_{1}$ is below $H_{2}$, hence there is inverted band ordering. Since the confining potential in the SDQW is symmetric, spin-orbit coupling is nonexistent, and thus the subbands are spin degenerate.

Similar to the case of BG, the band gap between the $H_{1}$ and $H_{2}$ subbands in the SDQW can be changed by means of an electric field, which is shown in Figure \ref{fig:2}(b). When an electric field is present, the spin degeneracy at $k_\parallel\neq 0$ is lifted, which is a demonstration of the Bychkov-Rashba effect. We might also note that a small indirect gap $\Delta$ emerges close to $k_\parallel=0$. Furthermore, the dispersion relations of the $H_1$ and $H_2$ subbands around this gap have a characteristic Mexican-hat shape \cite{He2019}. This is similar to bilayer graphene in the presence of a perpendicular electric field \cite{Min2007}, and is a consequence of mixing between the $H_1$ and $H_2$ subband states at finite $k_{\parallel}$. It results in an anticrossing between the two subbands, which in turn induces the opening of a gap close to $k_\parallel=0$. With increasing electric field, tunneling between the wells is enhanced, and as a consequence the anticrossing shifts to a larger $k_\parallel$. Since spin-orbit coupling lifts the spin degeneracy, both the width and the location of the energy gap become spin-dependent.

We found that $\Delta$ in the SDQW is of the order of a few meV for a relatively large electric field. For example, for $F_{el}=15$ {kV/cm}, we found $\Delta=12.8$ meV. Since the local extrema of the $H_1$ and $H_2$ subbands are located at $k_\parallel\neq 0$, $\rm{\Delta}$ is not a smooth function of the electric field \cite{Krishtopenko2016}. The electric field needed for a large gap could induce electric breakdown. But an increase of electric field has another effect: it shifts the band extrema and the energy gap to larger $k_\parallel$.

Instead of an electric field, we propose here that the asymmetry of the structural potential in the field-free case could be used to open an energy gap. Similar to the gap that arises due to the electric field in the SDQW system, the energy gap in the ADQW emerges between the $H_{1}$ and $H_{2}$ subbands. For example, when $w_1=6$ nm and $w_2=7.8$ nm, $\rm{\Delta}=8.6$ meV (see Figure \ref{fig:2}(c)). The electron-like $E_{1(2)}$ and heavy-hole-like $H_{1(2)}$ bands are not coupled at $k_\parallel=0$.  Furthermore, the states with different spins are not degenerate at $k_{\parallel}\neq 0$, as in the case of SDQW. But here the dispersion relations of the $H_1$ and $H_2$ subbands do not have the Mexican-hat shape. The states of the two subbands in the ADQW are differently localized, mostly in one quantum well, which brings about the gap opening. However, the gap in the ADQW takes place at $k_\parallel=0$. And only the $E_{1}$ and $H_{2}$ subbands are considerably spin split. This is a consequence of the coupling between the two subbands for $k_{\parallel}\neq 0$, which results in the anticrossing at $k_{x}=0.14\: \rm{nm}^{-1}$. At this point the states of the two subbands are strongly hybridized, and hence a small energy gap $\mathrm{\Delta_h}$ arises between them.

\begin{figure}
	\includegraphics{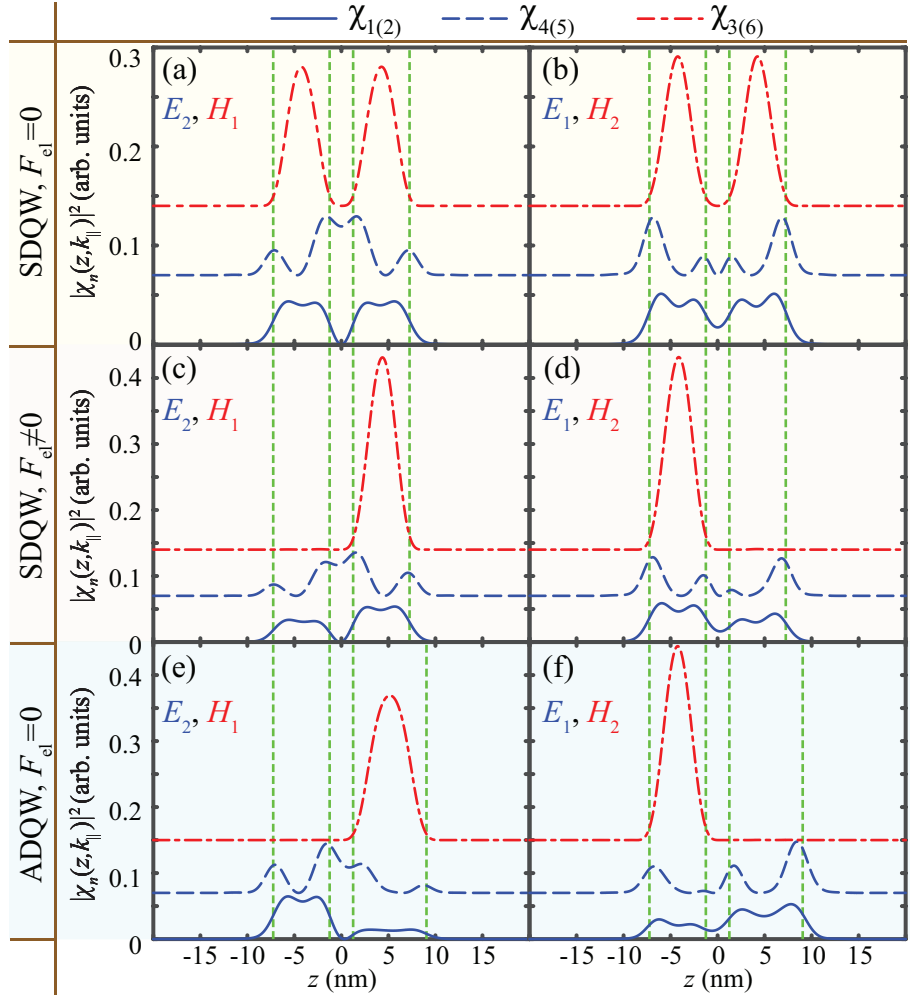}
	\caption{The probability density of the components of the $k_{||}=0$ subband states for (a,b) the SDQW with $F_{el}=0$, and (c,d) the SDQW for the electric field $F_{el}=15\: \rm{kV/cm}$, and (e,f) the ADQW for $F_{el}=0$. The probability density of the states of the $E_{2}$ and $H_{1}$ ($E_{1}$ and $H_{2}$) subbands are shown in the left (right) panel. For the sake of clearness, the shown probability densities are vertically separated by 0.07 (arb. units). Note that the probability densities of the Kramers degenerate counterparts coincide with each other.}
	\label{fig:3}
\end{figure}

In order to depict how the asymmetry of the potential affects the electron localization, the probability densities of the components of the $k_\parallel=0$ states of the four subbands are shown in Figure \ref{fig:3}. Here, the upper (middle) panel shows the probability densities in the SDQW in the absence (presence) of an external field, while the lower panel depicts the localization in the ADQW. The states of the $E_{2}$ and $H_{1}$ subbands are depicted in Figures \ref{fig:3}(a), (c), and (e), whereas the states of the $H_{2}$ and $E_{1}$ subbands are displayed in Figures \ref{fig:3}(b), (d), and (f). The electron-like subbands $E_{1}$ and $E_{2}$ are composed of the $\ket{\mathrm{\Gamma}_{6},\pm1/2}$ and $\ket{\mathrm{\Gamma}_{8},\pm1/2}$ zone-center states, while the hole-like $H_{1}$ and $H_{2}$ subband states are composed of the $\ket{\mathrm{\Gamma}_{8},\pm3/2}$ zone-center states. The states of the $H_{1}$ and $H_{2}$ subbands, and the $\ket{\mathrm{\Gamma}_{6},\pm1/2}$ component of the subbands $E_{1}$ and $E_{2}$ are mostly localized near the centers of the quantum wells. On the other hand, the $\ket{\mathrm{\Gamma}_{8},\pm1/2}$ components of $E_{1}$ and $E_{2}$ are localized close to the edges of the wells.

In the absence of an external electric field, the $H_1$ and $H_2$ states in the SDQW are degenerate at $k_\parallel=0$. These states are quite similarly localized in the two wells (see Figures \ref{fig:3}(a) and (b)), but we found that they have opposite parity. An electric field enhances the interwell tunneling, which shifts the $H_1$ and $H_2$ states toward either the left or the right quantum well (see Figures \ref{fig:3}(c) and (d)). Such an increase of the spatial separation between the two states leads to an energy gap opening between them. The electron and hole wavefunctions in the ADQW are obviously asymmetric (see Figures \ref{fig:3}(e) and (f)), and similar to the case of the SDQW and $F_{el}\neq 0$, both the envelope functions $\ket{\mathrm{\Gamma}_{6},\pm1/2}$ and $\ket{\mathrm{\Gamma}_{8},\pm1/2}$ are almost fully localized in one quantum well. Furthermore, the envelope function of the $\ket{\mathrm{\Gamma}_{8},\pm1/2}$ zone-center state is shifted close to the edges of the quantum wells. And the electron part of the multiband function in the higher (lower) energy $H_{1}$ ($H_{2}$) state is localized almost completely in the wider (narrower) well (see Figures \ref{fig:3}(e) and \ref{fig:3}(f)).

\begin{figure}
	\includegraphics{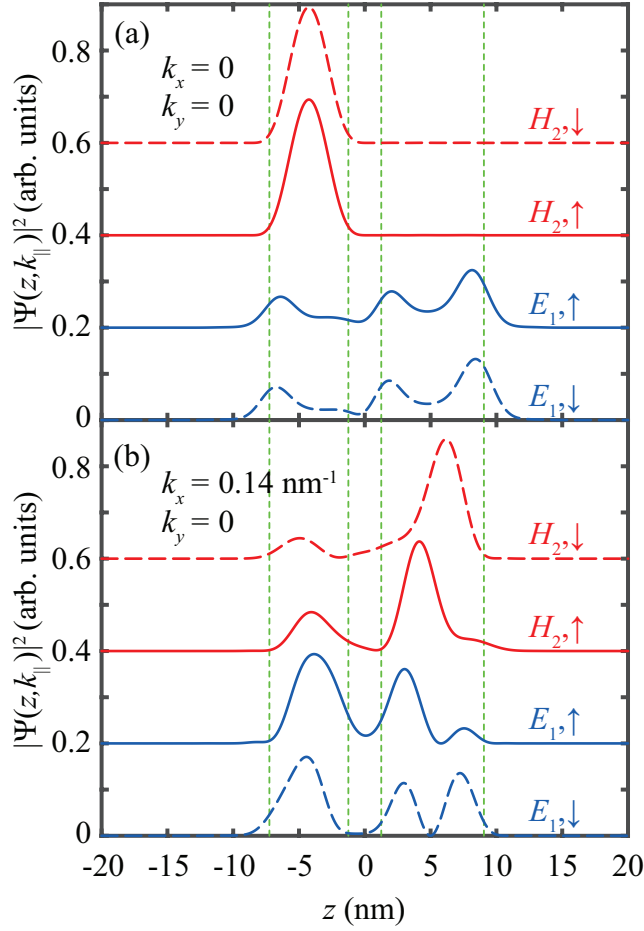}
	\caption{The total probability density at: (a) $(k_x,k_y)=(0,0)$ nm$^{-1}$ and (b) $(k_x,k_y)=(0.14,0)$ nm$^{-1}$ in the ADQW. Vertical (green) dotted lines denote layer boundaries. Solid (dashed) lines denote the spin-up (spin-down) states.  For the sake of clearness, different curves are vertically shifted by 0.2 (arb. units).}
	\label{fig:4}
\end{figure}

In order to explain the origin of the hybridization gap between the electron-like $E_1$ subband and the hole-like $H_2$ subband, we show in Figures \ref{fig:4}(a) and (b) the total probability density $|\Xi_{\bf k_\parallel}|^2=\sum_{n=1}^{8} |\chi_n ({\bf k_\parallel},z)|^2$ of these states at $(k_x,k_y)=(0,0)$ nm$^{-1}$ and $(k_x,k_y)=(0.14,0)$ nm$^{-1}$, respectively. At $(k_x,k_y)=(0,0)$ nm$^{-1}$ there is no coupling between the $E_1$ and $H_2$ subbands. Due to the spin degeneracy the total probability densities of the states of opposite spins are identical. As evident from Figure \ref{fig:4}(a), the $H_2$ subband state is almost fully localized in the narrower well, while the extrema of the $E_1$ envelope functions are located close to the interfaces of the wider well. When $k_\parallel$ increases, the $E_1$ and $H_2$ subbands start to couple. A fingerprint of this coupling is the observed spread of the probability densities of the states of the two subbands into separate wells. The overlap between the two states increases, implying they mutually hybridize. As could be inferred from Figure \ref{fig:4}(b), the largest overlap between the $E_1$ and $H_2$ states takes place at $k_x=0.14$ nm$^{-1}$. Here, the subbands anticross and an hybridized energy gap is opened. We might recall that the spin-orbit coupling due to the asymmetry of the confining potential in the ADQW affects the subbands, as shown in Figure \ref{fig:2}(c). Also, we note that the smallest energy gap $\Delta_h=3.9$ meV in the ADQW is opened between the spin-up $E_1$ and $H_2$ states.

\begin{figure}
		\includegraphics{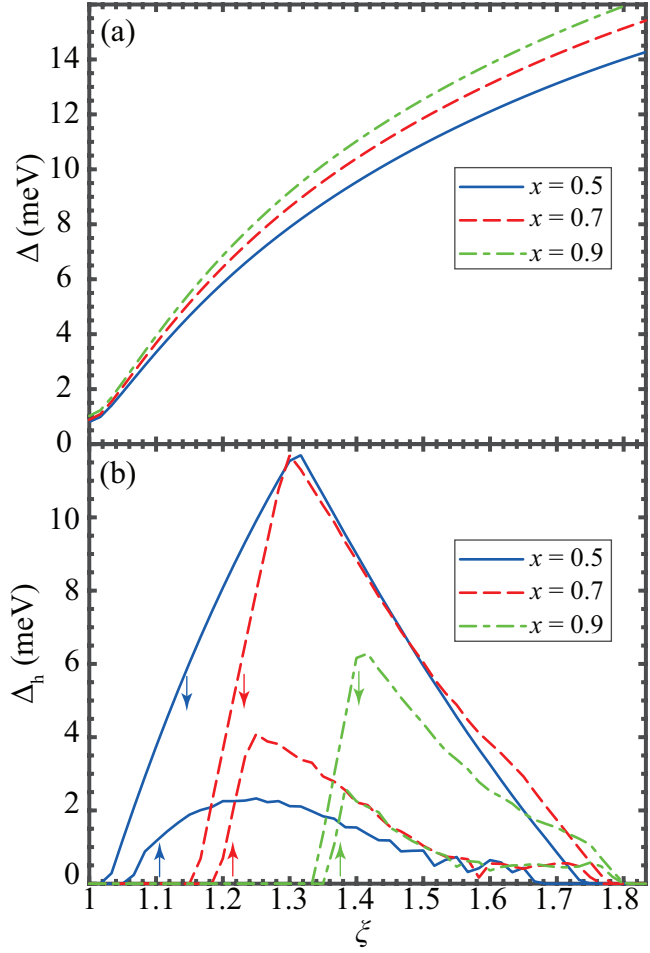}
	\caption{(a) The energy gap $\rm{\Delta}$ between the $H_{1}$ and $H_{2}$ subbands, and (b) the hybridization gap $\rm{\Delta_h}$  between the $E_{1}$ and $H_{2}$ subbands, as function of the relative ratio $\xi=w_{2}/w_{1}$ for the mole fractions $x\in\{0.5,0.7,0.9\}$ at $T=0$ K. The arrows in (b) denote the spin of the states.}
	\label{fig:5}
\end{figure}

The gap in the SDQW increases with electric field. And a similar conclusion might be {\it a priori} derived for the case of increasing structural asymmetry in the ADQW. It would affect the gaps which separates both different hole-like subbands and the electron- and the hole-like subands. For example, the width of only one quantum well in the ADQW could be varied, whereas the width of the other is kept constant. We choose $w_1=6$ nm, and vary $w_2$ in the range from 6 to 11 nm. $\Delta$ and $\Delta_h$ as function of the ratio $\xi=w_2/w_1$ for a few values of the mole fraction $x$ are shown in Figures \ref{fig:5}(a) and (b), respectively. We have previously found that the $H_1$ state at $k_\parallel=0$ is localized in the wider well, while the $H_2$ state is almost fully confined to the narrower quantum well. Thus, when $\xi$ increases, the difference between the energies of the two states increases. With other words, the energy gap $\Delta$ between the hole states increases, which is displayed in Figure \ref{fig:5}(a). Moreover, the energies of the subbands are strongly dependent on the mole fraction $x$, which is a way to tune the band gap in a double quantum well system \cite{Sengupta2013}. It relies on the increase of the band offset for the holes when $x$ increases, thus hole confinement is enhanced, and the interwell tunneling of the holes is reduced. Consequently, $\Delta$ increases.

Finally, we analyze the effects of the asymmetry and the mole fraction on the hybridization gap in the ADQW. Note that in the SDQW the $E_1$ and $H_2$ subbands become almost degenerate when $k_\parallel$ increases (see Figure \ref{fig:2}(a)). When one well is wider, the subbands whose wave functions are confined more in the narrower well are regularly ordered, whereas the subbands composed of the states that are more confined in the wider well are inverted. The $E_1$ and $H_2$ subbands split, and above a certain value of $\xi$, the hybridization gap opens at finite $k_\parallel$. As $\xi$ increases further, the higher energy light hole-like $H_2$ subband moves closer to the lower-energy electron-like $E_1$ subband. Due to the mixing, the gap between the $E_1$ and $H_2$ subbands decreases, and it eventually vanishes for large $\xi$. Moreover, when the mole fraction increases, the electrons and holes are more strongly confined, the anticrossing of the subbands becomes sharper, and thus $\Delta_h$ decreases.

\section{Summary and conclusion}
\label{sec:4}

The eigenstates in the asymmetric $\rm{HgTe/Cd_{x}Hg_{1-x}Te}$ double quantum wells where investigated by using the eight-band Kane $\bf{k}\cdot\bf{p}$ model. The energy gap that arises between the states of the total angular momentum $j=3/2$ is found to be affected by varying the ratio of the well widths ($\xi$). Furthermore, for a certain range of $\xi$, the electron-like and hole-like states become strongly hybridized, which leads to the opening of an hybridization gap at a nonzero in-plane wavevector. Our results show how the energy gaps in the asymmetric $\rm{HgTe/Cd_{x}Hg_{1-x}Te}$ double quantum wells can be tuned by means of either varying the well widths or changing the composition of the ${\rm Cd}_{x}{\rm Hg}_{1-x}{\rm Te}$ interwell barrier. We found that the hybridization gap could be as large as 12 meV in the absence of a gate voltage, and is comparable to the energy gap which appears in the symmetric system when a large electric field of 15 kV/cm is applied. Also, the subbands exhibit inverted ordering, thus when properly designed the analyzed asymmetric system should exhibit topological effects similar to other $\rm{HgTe/Cd_{x}Hg_{1-x}Te}$ quantum-well structures. As a matter of fact, properly design asymmetric quantum well-based systems should exhibit the spin-polarized counter-propagating edge channels, which is a necessary ingredient to establish applications of the ADQW's in state-of-the-art spintronic devices.

\begin{acknowledgments}
	The research was funded by the Ministry of Education, Science and Technological Development of the Republic of Serbia, and the Flemish Science Foundation (FWO-Vl).
\end{acknowledgments}

\section*{Data availability statement}

The data that support the findings of this study are available from the corresponding author upon reasonable request.

\appendix
\section{}
\label{app:A}

The $\bf{k}\cdot\bf{p}$ theory is an empirical method for calculating the band structure of bulk semiconductors \cite{Kane1956}. In this approach, an unknown wave function is represented within the basis of Bloch functions in the center of the Brillouin zone. HgTe has an inverted band structure, and degenerate hole-like bands at the $\Gamma$ point \cite{Winkler2003, Novik2005}. Its zinc blende crystalline structure consists of heavy Hg and Te atoms, in which relativistic effects are enhanced. To account for the spin-orbit coupling, as well as the mixing between the conduction and valence band states, we use the eight-band Kane model with the basis set
\begin{eqnarray}
	u_{1}=\ket{\mathrm{\Gamma}_{6},+1/2}=S\uparrow,
	\nonumber\\
	u_{2}=\ket{\mathrm{\Gamma}_{6},-1/2}=S\downarrow,
	\nonumber\\
	u_{3}=\ket{\mathrm{\Gamma}_{8},+3/2}=\frac{1}{\sqrt{2}} (X+iY)\uparrow,
	\nonumber\\
	u_{4}=\ket{\mathrm{\Gamma}_{8},+1/2}=\frac{1}{\sqrt{6}} [(X+iY)\downarrow-2Z\uparrow],
	\nonumber\\
	u_{5}=\ket{\mathrm{\Gamma}_{8},-1/2}=-\frac{1}{\sqrt{6}} [(X-iY)\uparrow+2Z\downarrow],
	\nonumber\\
	u_{6}=\ket{\mathrm{\Gamma}_{8},-3/2}=-\frac{1}{\sqrt{2}} (X-iY)\downarrow,
	\nonumber\\
	u_{7}=\ket{\mathrm{\Gamma}_{7},+1/2}=\frac{1}{\sqrt{3}} [(X+iY)\downarrow+Z\uparrow],
	\nonumber\\
	u_{8}=\ket{\mathrm{\Gamma}_{7},-1/2}=\frac{1}{\sqrt{3}} [(X-iY)\uparrow-Z\downarrow].
	\label{eq:6}
\end{eqnarray}

This model can be efficiently extended to a multi-layered heterostructures using the same basis set for all constituent materials and by replacing the Bloch phase factor by an envelope function respecting the appropriate symmetrization rules in order to avoid appearance of spurious solutions \cite{Burt1999}. The Hamiltonian for quantum wells grown along the [001] direction has the form:\cite{Novik2005}
\begin{eqnarray}
	H({\mathbf{k}})=	
	\left[\matrix{
		T & 0 & -\frac{1}{\sqrt{2}}Pk_{+} & \sqrt{\frac{2}{3}}Pk_{z} & \frac{1}{\sqrt{6}}Pk_{-} & 0 & -\frac{1}{\sqrt{3}}Pk_{z} & -\frac{1}{\sqrt{3}}Pk_{-} \cr
		0 & T & 0 & -\frac{1}{\sqrt{6}}Pk_{+} & \sqrt{\frac{2}{3}}Pk_{z} & \frac{1}{\sqrt{2}}Pk_{-} & -\frac{1}{\sqrt{3}}Pk_{+} & \frac{1}{\sqrt{3}}Pk_{z} \cr
		-\frac{1}{\sqrt{2}}k_{-}P & 0 & U+V & -\bar{S}_{-} & R & 0 & \frac{1}{\sqrt{2}}\bar{S}_{-} & -\sqrt{2}R \cr
		\sqrt{\frac{2}{3}}k_{z}P & -\frac{1}{\sqrt{6}}k_{-}P & -\bar{S}_{-}^{\dagger} & U-V & C & R & \sqrt{2}V & -\sqrt{\frac{3}{2}}\tilde{S}_{-} \cr
		\frac{1}{\sqrt{6}}k_{+}P & \sqrt{\frac{2}{3}}k_{z}P & R^{\dagger} & C^{\dagger} & U-V & \bar{S}_{+}^{\dagger} & -\sqrt{\frac{3}{2}}\tilde{S}_{+} & -\sqrt{2}V \cr
		0 & \frac{1}{\sqrt{2}}k_{+}P & 0 & R^{\dagger} & \bar{S}_{+} & U+V & \sqrt{2}R^{\dagger} & \frac{1}{\sqrt{2}}\bar{S}_{+} \cr
		-\frac{1}{\sqrt{3}}k_{z}P & -\frac{1}{\sqrt{3}}k_{-}P & \frac{1}{\sqrt{2}}\bar{S}_{-}^{\dagger} & \sqrt{2}V & -\sqrt{\frac{3}{2}}\tilde{S}_{+}^{\dagger} & \sqrt{2}R & U-\rm{\Delta} & C \cr
		-\frac{1}{\sqrt{3}}k_{+}P & \frac{1}{\sqrt{3}}k_{z}P & -\sqrt{2}R^{\dagger} & -\sqrt{\frac{3}{2}}\tilde{S}_{-}^{\dagger} & -\sqrt{2}V & \frac{1}{\sqrt{2}}\bar{S}_{+}^{\dagger} & C^{\dagger} & U-\rm{\Delta}} \right],\cr\cr \label{eq:7}
\end{eqnarray}
where
\begin{eqnarray}
	T=E_{c}+\theta[(2F+1)k_{||}^2+k_{z}(2F+1)k_{z})],
	\nonumber\\
	U=E_{v}-\theta(\gamma_{1}k_{||}^2+k_{z}\gamma_{1}k_{z}),
	\nonumber\\
	V=-\theta(\gamma_{2}k_{||}^2-2k_{z}\gamma_{2}k_{z}),
	\nonumber\\
	R=-\theta(\sqrt{3}\mu k_{+}^2-\sqrt{3}\bar{\gamma}k_{-}^2),
	\nonumber\\
	\bar{S}_{\pm}=-\theta\sqrt{3}k_{\pm}(\{\gamma_{3},k_{z}\}+[\kappa,k_{z}]),
	\nonumber\\
	\tilde{S}_{\pm}=-\theta\sqrt{3}k_{\pm}(\{\gamma_{3},k_{z}\}-\frac{1}{3}[\kappa,k_{z}]),
	\nonumber\\
	C=2\theta k_{-}[\kappa,k_{z}],
	\nonumber\\
	k_{||}^2 = k_{x}^2+k_{y}^2,
	\nonumber\\
	k_{\pm}= k_{x}\pm ik_{y},
	\nonumber\\
	k_{z}=-i\frac{\partial}{\partial z}.
	\label{eq:8}
\end{eqnarray}
Here, $\theta=\hbar^{2}/2m_{0}$, $k_{x}$ and $k_{y}$ are good quantum numbers, while $k_{z}$ is an operator. $\gamma_{1}$, $\gamma_{2}$ and $\gamma_{3}$ are the Luttinger parameters, $\kappa$, and $F$ describe the coupling to the remote bands, while $\bar{\gamma}=(\gamma_{3}+\gamma_{2})/2$ and $\mu=(\gamma_{3}-\gamma_{2})/2$. The values of the band-structure parameters in CdTe and HgTe at $T=0 \rm{K}$ are given in Table \ref{tab:2}.

\begin{table}
	\caption{The parameters of the band structure of CdTe and HgTe at $T=0 \rm{K}$ \cite{Novik2005}.}
	\label{tab:2}
	\begin{ruledtabular}	
		\begin{tabular}{lllllllll}
			Material & $\gamma_{1}$ & $\gamma_{2}$ & $\gamma_{3}$ & $\kappa$ & $F$ & $\rm{\Delta} (\rm{eV})$ & $E_{g} (\rm{eV})$ & $2m_{0}P^{2}/\hbar^{2} (\rm{eV})$  \\
			\hline
			CdTe & 1.47 & -0.28 & 0.03 & -1.31 & -0.09 & 0.91 & 1.606 & 18.8\\
			HgTe & 4.1 & 0.5 & 1.3 & -0.4 & 0 & 1.08 & -0.303 & 18.8 \\
		\end{tabular}
	\end{ruledtabular}		
\end{table}


\end{document}